\documentclass[useAMS,usenatbib]{mn2e}

\usepackage[utf8]{inputenc}
\usepackage{graphicx}
\usepackage{amsmath}
\usepackage{wasysym}
\usepackage{url}

\title[Timing of Eclipsing Binary Stars from the ASAS Catalogue]
{Radio Pulsar Style Timing of Eclipsing Binary Stars from the ASAS Catalogue}
\author[S. K. Koz\l owski, M. Konacki and P. Sybilski]{S. K. Koz\l owski$^{1,2}$\thanks{E-mail:
stan@ncac.torun.pl}, M. Konacki$^{1,2}$\thanks{E-mail:
maciej@ncac.torun.pl} and P. Sybilski$^{2}$\thanks{E-mail:
sybilski@ncac.torun.pl}\\
$^{1}$Adam Mickiewicz University, Astronomical Observatory, Pozna\'{n}, Poland\\
$^{2}$Nicolaus Copernicus Astronomical Center, Toru\'{n}, Poland }
\begin{document}

\date{Received...}

\pagerange{\pageref{firstpage}--\pageref{lastpage}} \pubyear{2010}

\maketitle

\label{firstpage}

\begin{abstract}
The Light-Time Effect (LTE) is observed whenever the distance between the observer and any kind 
of periodic event changes in time. The usual cause of this distance change 
is the reflex motion about the system's barycenter due to the gravitational influence of one 
or more additional bodies. We analyze 5032 eclipsing contact (EC) and detached
(ED) binaries from the All Sky Automated Survey (ASAS) catalogue to detect variations in
the times of eclipses which possible can be due to the LTE effect. 
To this end we use an approach known from the radio pulsar timing where a template 
radio pulse of a pulsar is used as a reference to measure the times of arrivals
of the collected pulses. In our analysis as a template for a
photometric time series from ASAS, we use a best-fitting trigonometric series 
representing the light curve of a given EC or ED. Subsequently, an O--C diagram 
is built by comparing the template light curve with light curves obtained
from subsets of a given time series.  Most of the variations we detected in
O--Cs correspond to a linear period change. Three show evidence of more than one complete 
LTE-orbit. For these objects we obtained preliminary orbital solutions.
Our results demonstrate that the timing analysis employed in radio pulsar
timing can be effectively used to study large data sets from photometric surveys.
\end{abstract}


\section{Introduction}
The fact that the velocity of light is finite was not obvious till 1676 when Olaus Roemer 
carried out precise measurements of the times of eclipses of Jovian moons. He noted that 
Io eclipses were "early" before opposition and "late" after opposition when compared to 
the \textit{Ephemerides Bononiensis Mediceorum Siderum}, a~work by Cassini published 
in 1668. It includes tables of times of eclipses of Jovian moons which were used to 
determine the differential longitude by simultaneous observations of the same eclipse 
from two places. Roemer's conclusion, though not a~quantitative one, became a~great
discovery contradicting the Aristotelean thought. He provided scientists with
the basics of the O--C (observed minus calculated) procedure \citep{Sterken05Roemer} 
and was the first one to analyze the effects caused by finite light speed, hereafter 
called the light time effect (LTE). 

Below we analyze the photometric data from the All Sky Automated Survey, 
\citep[ASAS;][]{Pojmanski02}. In \S{2} we present our method for analyzing 
the timing variations. In \S{3} we show the outcome of applying our
approach to the photometric series of 5032 eclipsing contact (EC) and detached
(ED) binaries from ASAS. In \S{4} we discuss several interesting cases of
most likely the LTE effect due to companions to the analyzed systems and 
conclude in \S{5}.

\section[]{Automated timing of eclipsing binaries}

\begin{figure}
	\includegraphics[width=\columnwidth]{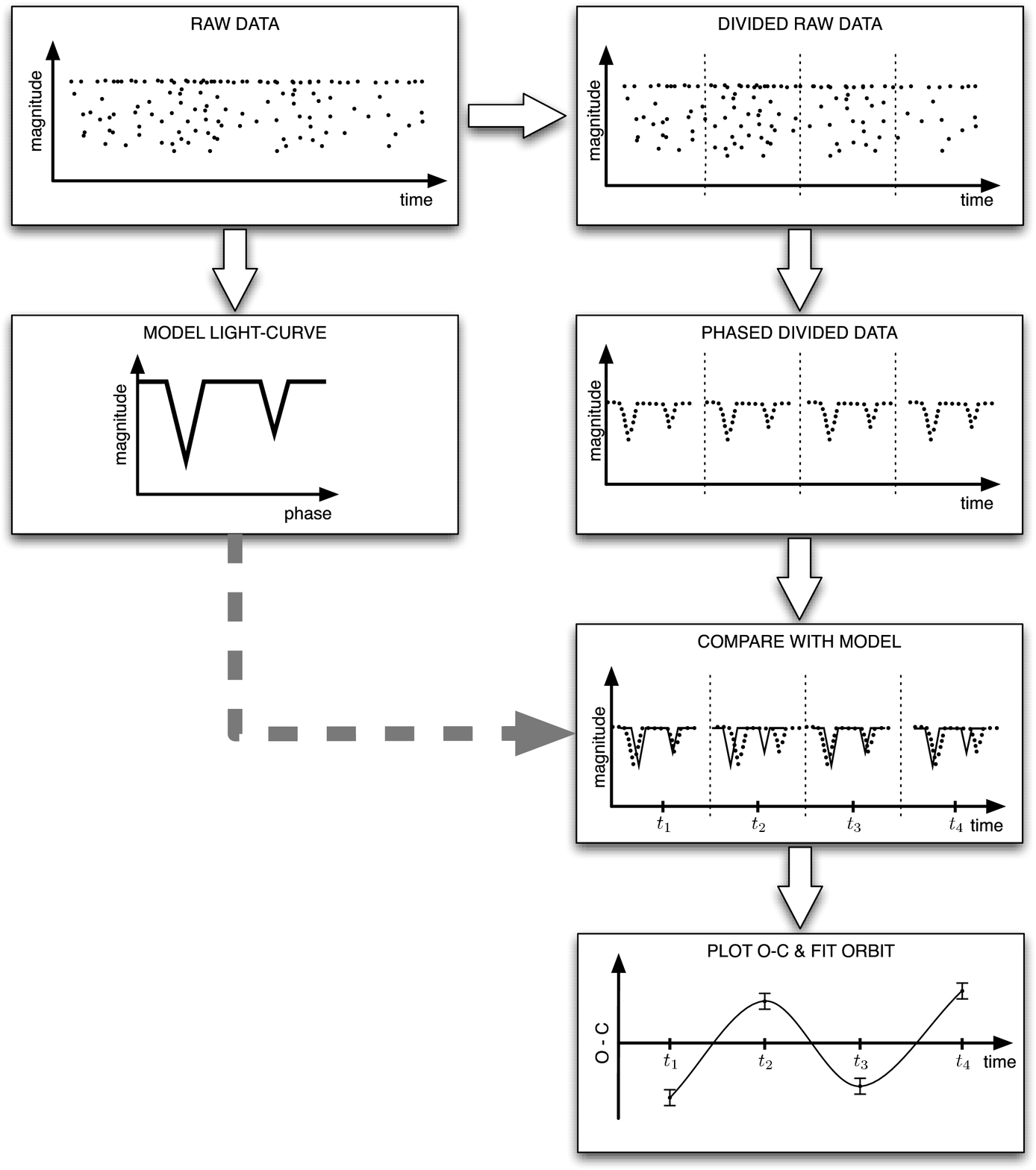}
	\caption{A block diagram describing our eclipse timing procedure.}
	\label{fig:AlgorithmDiagram}
\end{figure}

Our basic concept of detecting timing variations in photometry of ED/EC 
in an automated way is based on the method used in radio pulsar timing.  
It consists of six steps which are shown as a~block diagram in 
Fig. \ref{fig:AlgorithmDiagram}. 

\begin{enumerate}
\item\textbf{Get a raw data set.}
It is assumed that a raw data set consists of $n$ magnitudes (or fluxes), their 
errors and the times they were recorded. 
\item\textbf{Prepare a template light-curve model.}
A raw data set is phased with the known period and the parameters of a
template light-curve model are calculated using the least squares method. 
At this stage the period can be improved or corrected during the fitting 
process. This is often necessary for the ASAS data.
\item\textbf{Divide a raw data set into subsets.}
A raw data set is divided into $M$ intervals. The intervals can be equal in terms of the 
number of data points they include or the time that they span. The second variant 
was chosen in this paper. The intervals can be overlapping or have a non-repeating 
content. When implementing the first method, appropriate corrections must be applied 
when calculating the final formal errors due to a multiple usage of the same data 
points.
\item\textbf{Phase data in each interval.}
%Data points are phased separately in each interval. This way a~\textit{local light-curve} 
%is created and the mid-time of each interval is associated with it. 
Data points are phased separately in each interval using the new period value calculated during the creation of the model. The zeropoint is retained for each interval. This way a~\textit{local light-curve} 
is created and the mid-time of each interval is associated with it.
\item\textbf{Compare with the template light-curve model.}
The most important step in this procedure is the comparison of the \textit{local light-curves} 
with the template light-curve model. A this stage a one-parameter least squares fit is 
performed in order to find the time shift between the two light curves.
\item\textbf{Plot an O--C diagram, fit an LTE orbit.}
Finally, the collected O--C values can be plotted against time and, if possible, an orbit can be fitted. 
\end {enumerate}

\subsection{A template light-curve model}

\begin{figure}
	\includegraphics[width=\columnwidth]{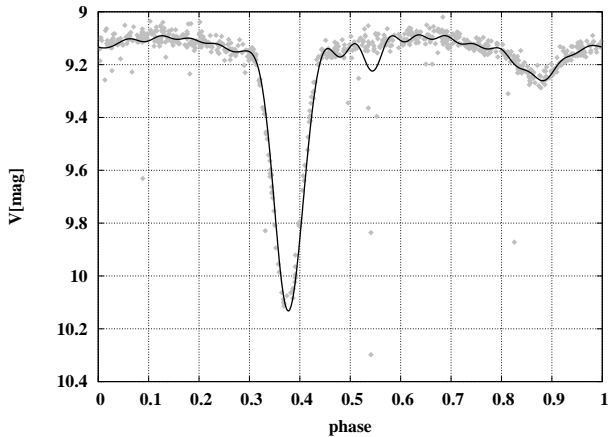}
	\caption{An example template light-curve model (solid line) calculated for ASAS 023539-4504.2 
raw data. Note that the solid curve approximates the primary minimum very well despite 
oscillations visible between eclipses.}
	\label{fig:harmonic18}
\end{figure}

We have tested two representations for a template model light curve --- 
a polynomial model and a harmonic model. We have decided to use 
a harmonic model. Such an approach, apart from providing a~good model of the 
input data, enables us to conveniently adjust the initial period of a binary. 
The model is based on a Fourier series and involves fitting a~trigonometric series 
to a raw photometric data set:
\begin{equation}
f(t) = \frac{a_0}{2} + \sum\limits_{l=1}^{N}\left[ a_l\cos\left(\frac{2\pi }{T }lt\right) 
+b_l\sin\left(\frac{2\pi }{T}lt\right)\right].
\label{eqn:FourierSeriesModel}
\end{equation}
$N$ has to be chosen so that the resulting light-curve model defined by the coefficients 
$a_0\ldots a_N$, $b_1\ldots b_N$ and the period $T$ approximate the raw data as good as 
possible. Theoretically, the more harmonics are used (big $N$), the better the approximation. 
$N$ has an upper limit though.  Obviously, as a least squares algorithm is used to fit $f(t)$,  
the number of parameters cannot exceed the number of data points used in a fitted, i.e. 
$m=2N+2<n$. Moreover, if $m$ is close to $n$, the fit starts to approximate the data noise. 
This is not a~desired effect  In our analysis we used $N=18$ which was found to be the best for the types of curves analyzed.  An example
template model along with the original light-curve are presented in Fig. \ref{fig:harmonic18}. Though it might seem that a lower value of $N$  would make the model less sensitive to erroneous data points, at the same time the real eclipses (especially deep and short ones) would not have been modeled well enough. Having the primary eclipse well modeled is crucial when searching for time shifts between the model and the local light curves. If one would have high quality photometry it would then be good to optimize the procedure and select N individually for each object.
In order to carry out the least-squares fitting we used the Levenberg-Marquardt algorithm and its 
Minpack\footnote{\texttt{http://netlib.org/minpack/}} implementation. 

%Let us
%also note that in radio pulsar timing an FFT is used to represent pulsar
%pulses. A trigonometric series can be considered as an FT for unevenly
%sampled data. 

\subsection{Calculating O--C}
Having the light curve model in form of $T$ and $a_0\ldots a_N$, $b_1\ldots b_N$ coefficients, it can be compared with local light curves in each interval. This is done by fixing the parameters describing the model in Eqn. \eqref{eqn:FourierSeriesModel} and slightly modifying the formula by introducing a time shift parameter $t_O$:
\begin{equation}
\begin{split}
f_O(t) = \frac{a_0}{2} + \sum\limits_{l=1}^{N}& \left[ a_l\cos\left(\frac{2\pi }{T }l(t-t_O)\right)\right. \\
&+\left. b_l\sin\left(\frac{2\pi }{T}l(t-t_O)\right)\right].
\label{eqn:FourierSeriesModel2}
\end{split}
\end{equation}
Then, using least squares, $f_O(t)$ is fitted to local light curves with $t_O$ as the only parameter. Effectively, $t_O$ is the value of O--C at the given point in time. Collecting these for all intervals allows one to obtain a O--C diagram. Since Eqn. \eqref{eqn:FourierSeriesModel2} is fitted using the Levenberg-Marquardt as  before, the formal errors of the obtained O--C values are derived from the covariance matrix, which, in this case is a one-element matrix  due to the fact that the fit has only one parameter.

\subsection{Detection Criterion}
Inspecting visually every single O--C diagram is not practical, hence in order to
find binaries with significant timing variations we use the following
timing activity parameter $r$
\begin{equation}
r = \frac{S}{A}>r_{min},
\label{eqn:criterion}
\end{equation}
where $S$ is the standard deviation of the O--C values and $A$ is the average error ($\Delta(O-C$))
of these values:
\begin{equation}
A = \frac{1}{M}\sum\limits_{l=1}^{M}\Delta(O-C)_l.
\end{equation}
Objects having $r$ greater than a certain $r_{min}$ are considered 
interesting. We applied the above criterion on data sets divided into 
5-, 6-, 7- and 8-intervals. If an object passes the criterion at least
once, it is considered interesting. Such objects are finally inspected
visually. 

\section{Timing variations of ED and EC binaries from the ASAS catalogue}

\textit{The ASAS Catalogue of Variable Stars} (ACVS) is publicly available for download 
from the ASAS Project homepage\footnote{\texttt{http://www.astrouw.edu.pl/asas/}}. 
We used its version 1.1 in this paper. The catalogue consists of 50124 objects 
showing variability in brightness. Among them 2761 are uniquely classified as eclipsing 
contact (EC) binaries and 2271 as eclipsing detached (ED) binaries. Based on the original 
ACVS index file, we created two subindex files: one containing only EC binaries and
a~second one containing only ED binaries. This reduced the number of investigated light-curves 
to 5032. In the analysis only an A-rated\footnote{The ACVS rates the quality of each brightness measurement on a scale form A to D with A being the best and D the worst quality.} photometry was taken into account.
The results for ED are collected in Table \ref{tab:ED}. It lists 29
binaries. For this table we used $r_{min}=2.0$ . The results for EC
are collected in Table \ref{tab:EC}. It contains 44 objects. In this
case we used $r_{min}=2.5$. The most frequent variation in O--C in both
cases has a parabolic shape indicating a linear change in period.
The $r_{min}$ threshold has been adjusted individually for each type of curves (ED and EC). In case of contact binaries the eclipses are generally shallower and less sharp than in case of detached binaries - this results in different sensitivities of the algorithm in the two cases.

\begin{table*}
\small\centering
\begin{tabular}{cccccccc}
\hline
ASAS ID & $P$[d] (ASAS) & $P$[d] (corrected) & $r_5$ & $r_6$ & $r_7$ & $r_8$ & Other ID\\
\hline
\hline
034746-0836.7	&	2.8764	&	2.8768183	&	3.58	&	1.66	&	2.11	&	2.15	&	CD~Eri              \\
050205-2842.8	&	3.3023	&	3.3024868	&	2.01	&	2.73	&	1.73	&	1.91	&	-                   \\
053727-7752.3	&	0.99158	&	0.9915804	&	3.14	&	2.74	&	2.25	&	1.92	&	-                   \\
070825-4433.2	&	1.8519	&	1.8518267	&	2.17	&	1.97	&	1.15	&	2.24	&	-                   \\
071021-3324.6	&	1.657725	&	1.6577068	&	2.07	&	2.18	&	1.15	&	0.89	&	CI~Pup              \\
090039-4739.8	&	4.4045	&	4.4047483	&	2.09	&	2.16	&	1.95	&	1.56	&	-                   \\
092456-3337.2	&	1.44643	&	1.4464050	&	2.42	&	2.23	&	2.45	&	1.15	&	SV~Pyx              \\
094542-4913.5	&	1.552517	&	1.5525371	&	3.00	&	3.38	&	3.27	&	1.86	&	DU~Vel              \\
111915-1949.7	&	2.3409	&	2.3410108	&	3.80	&	3.71	&	3.10	&	2.33	&	RV~Crt              \\
121103-5040.3	&	1.9508	&	1.9507456	&	2.28	&	2.05	&	1.87	&	1.58	&	NSV05487            \\
121158-5050.7	&	1.135683	&	1.1356929	&	2.16	&	2.03	&	1.63	&	2.15	&	NSV05497            \\
130155-5040.7	&	3.511604	&	3.5117345	&	2.96	&	2.03	&	1.88	&	1.94	&	NSV06061            \\
130856-7437.6	&	1.479905	&	1.4799275	&	2.44	&	2.49	&	2.75	&	2.21	&	-                   \\
132107-1936.4	&	3.042031	&	3.0420286	&	2.71	&	2.45	&	1.48	&	1.32	&	-                   \\
132402-6345.9	&	1.737062	&	1.7370648	&	5.64	&	5.36	&	4.32	&	4.09	&	-                   \\
132538-2025.0	&	0.47849	&	0.4784917	&	2.30	&	1.99	&	1.90	&	1.86	&	-                   \\
141035-4546.8	&	0.988708	&	0.9887115	&	2.04	&	1.86	&	1.85	&	1.42	&	-                   \\
143636-5124.8	&	1.45313	&	1.4530916	&	1.84	&	1.71	&	2.01	&	1.68	&	DT~Lup              \\
154645-2307.5	&	1.281968	&	1.2819585	&	2.87	&	2.43	&	2.00	&	2.36	&	-                   \\
161628-0658.7	&	2.446109	&	2.4461205	&	3.03	&	2.04	&	1.99	&	1.59	&	SW~Oph              \\
171519-3639.0	&	2.41344	&	2.4134785	&	2.05	&	1.50	&	1.47	&	1.99	&	V0467~Sco           \\
174303-3222.3	&	2.192577	&	2.1925951	&	2.93	&	1.63	&	1.65	&	1.33	&	V0496~Sco           \\
191350+1109.8	&	0.334838	&	0.3348409	&	1.76	&	1.58	&	2.04	&	2.02	&	-                   \\
193840-4500.6	&	1.35187	&	1.3518939	&	2.73	&	2.97	&	2.96	&	1.69	&	V0795~Sgr           \\
200048-2833.4	&	1.665189	&	1.6651979	&	3.87	&	2.57	&	2.47	&	1.22	&	V1173~Sgr           \\
205101-6341.5	&	2.5442	&	2.5441900	&	2.68	&	1.23	&	1.01	&	1.16	&	BT~Pav              \\
213148-4502.7	&	1.880496	&	1.8805027	&	1.57	&	2.23	&	1.94	&	1.67	&	U~Gru               \\
215704-5606.0	&	0.454887	&	0.4548896	&	1.90	&	2.00	&	1.66	&	1.05	&	-                   \\
223621-1116.4	&	1.62854	&	1.6285257	&	2.70	&	1.36	&	1.71	&	2.02	&	-                   \\
\hline
\end{tabular}
\caption[LTE search results: ASAS ED objects.]{LTE search results: ASAS ED objects. 
The table lists the object's ID, the period value from the ACVS catalogue, a~corrected 
period obtained during reference model computation. Four $r$ values for 5-, 6-, 7- and 
8-interval runs are given. The last column contains other IDs, if any.}
\label{tab:ED}
\end{table*}

\begin{figure}
\centering
\includegraphics[width=\columnwidth]{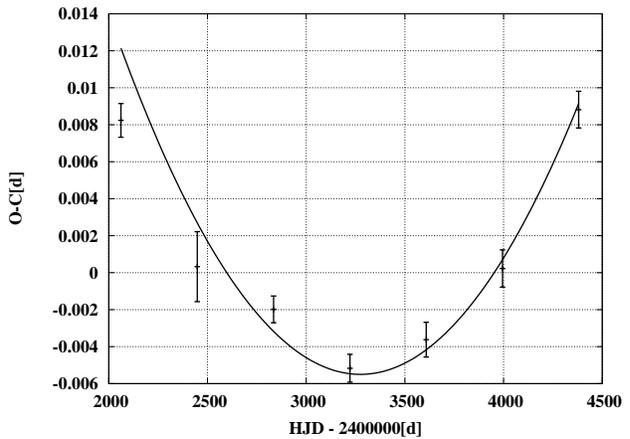}
\caption[ASAS 060557-5342.9: comparative plot.]{ASAS 060557-5342.9's
timing variations. The parabola corresponds to $\dot{P}$ from 
\cite{Pilecki07}.}
\label{fig:pil1}
\end{figure}

\begin{figure}
\centering
\includegraphics[width=\columnwidth]{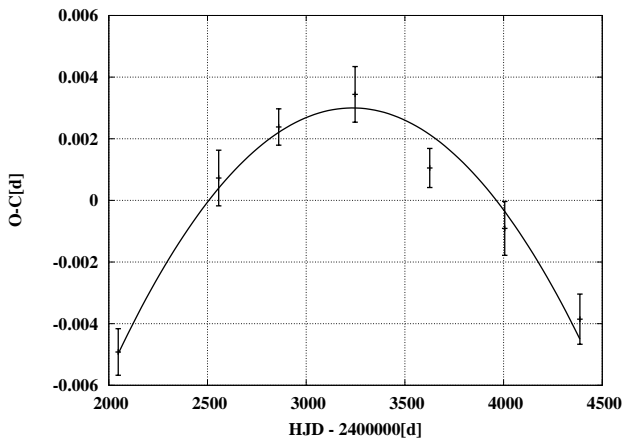}
\caption[ASAS 160302-3749.4: comparative plot.]{ASAS 160302-3749.4's
timing variations. The parabola corresponds to $\dot{P}$ from 
\cite{Pilecki07}.}
\label{fig:pil5}
\end{figure}

\begin{figure*}
\centering
\includegraphics[width=\textwidth]{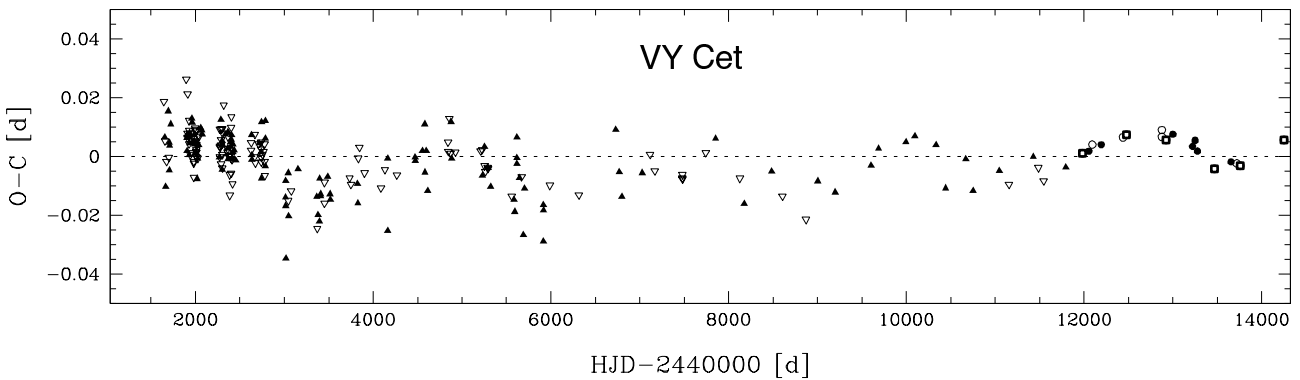}
\caption[Updated VY Cet: O--C diagram.]{The timing variations of VY Cet.
This O--C diagram is from \cite{Pilecki07}. Our timing variations
are denoted with squares.}
\label{fig:vycet_updated}
\end{figure*}

Recently \cite{Pilecki07} have found EC binaries with high period change rates (HPCR) in the ASAS data. 
They have published a~list of 31 stars exhibiting large $\dot{P}$. 10 out of 44 objects listed in 
Tab. \ref{tab:EC} have been detected by \cite{Pilecki07}. Of the remaining 21 HPCR objects, 12 were 
classified as other than EC hence they were not analyzed by us. Finally, 9 objects did not satisfy 
our criterion. In order to verify that the results obtained with our proposed algorithm are 
consistent with \cite{Pilecki07}, we compared objects with high $\dot{P}$ from the HPCR list 
with our timing measurements. This is demonstrated in Figures \ref{fig:pil1}-\ref{fig:pil5}  
by plotting the linear $\dot{P}$ (i.e. parabolic in O--C) trend from
\cite{Pilecki07} together with our timing measurements. We have also
extended the O--C diagram from \cite{Pilecki07} for VY Cet, a contact binary system with the well known
O--C variations. \cite{Qian} studied this object and found the period of the third body  to be 7.3 years 
with a~minimum mass of $0.62M_{\astrosun}$. Figure \ref{fig:vycet_updated} shows the original O--C 
diagram with our results overplotted.  

\begin{table*}\footnotesize
\begin{tabular}{cccccccc}
\hline
ASAS ID & $P$[d] (ASAS) & $P$[d] (corrected) & $r_5$ & $r_6$ & $r_7$ & $r_8$ & Other ID\\
\hline
\hline
002449-2744.3	&	0.31367	&	0.3136602	&	3.10	&	2.71	&	2.35	&	2.66	&	-                   \\
004240-2956.7	&	0.301682	&	0.3016864	&	3.49	&	3.15	&	2.29	&	3.25	&	-                   \\
030313-2036.9	&	0.334978	&	0.3349778	&	2.69	&	2.63	&	1.38	&	2.19	&	TU~Eri              \\
030315-2311.2	&	0.4566	&	0.4566055	&	2.85	&	2.56	&	2.26	&	1.99	&	-                   \\
030617-6812.5	&	0.41612	&	0.4161200	&	4.35	&	4.23	&	3.77	&	3.97	&	NSV01054            \\
032812-2503.5	&	0.315501	&	0.3155051	&	2.64	&	2.63	&	1.81	&	2.22	&	-                   \\
043046-4813.9	&	0.35714	&	0.3571467	&	3.59	&	3.56	&	2.19	&	3.06	&	-                   \\
050922-1932.5	&	0.270842	&	0.2708428	&	3.04	&	2.58	&	1.83	&	2.18	&	-                   \\
051114-0833.4	&	0.4234	&	0.4234030	&	4.47	&	3.79	&	3.30	&	3.21	&	ER~Ori              \\
052313-0907.7	&	0.40198	&	0.4019834	&	4.40	&	3.69	&	2.21	&	2.38	&	-                   \\
054000-6828.7	&	0.36222	&	0.3622215	&	2.79	&	2.76	&	2.31	&	3.03	&	ASAS~054000-6828.6  \\
055501-7241.6	&	0.343841	&	0.3438400	&	3.21	&	2.96	&	2.96	&	3.02	&	BV~435              \\
060557-5342.9	&	0.46363	&	0.4636373	&	4.61	&	4.45	&	3.48	&	3.03	&	-                   \\
061654-4326.4	&	0.504735	&	0.5047355	&	3.98	&	3.51	&	2.14	&	2.88	&	-                   \\
062254-7502.0	&	0.257704	&	0.2577062	&	5.18	&	5.28	&	5.24	&	4.91	&	-                   \\
064047-8815.4	&	0.43863	&	0.4386210	&	4.31	&	3.62	&	3.06	&	3.56	&	-                   \\
065232-2533.5	&	0.418634	&	0.4186402	&	2.79	&	2.63	&	1.98	&	2.13	&	-                   \\
070225-2845.8	&	0.462724	&	0.4627283	&	5.89	&	4.82	&	4.49	&	4.36	&	-                   \\
070232-5214.6	&	0.407338	&	0.4627283	&	5.89	&	4.82	&	4.49	&	4.36	&	-                   \\
070943-0702.3	&	0.501369	&	0.5013665	&	3.58	&	3.05	&	2.28	&	2.28	&	-                   \\
071727-4007.7	&	0.320267	&	0.3202645	&	4.46	&	4.55	&	3.09	&	3.62	&	GZ~Pup              \\
074308-1915.5	&	0.403302	&	0.4033032	&	3.78	&	3.84	&	2.91	&	3.22	&	-                   \\
075809-4648.5	&	0.390387	&	0.3903834	&	5.01	&	5.38	&	3.81	&	4.47	&	NSV03836            \\
082030-4326.7	&	0.37078	&	0.3707788	&	3.02	&	3.00	&	2.33	&	2.22	&	-                   \\
082456-4833.6	&	0.364879	&	0.3648710	&	6.88	&	5.69	&	4.86	&	3.67	&	-                   \\
083139-4227.5	&	0.302677	&	0.3026776	&	3.00	&	2.75	&	3.28	&	2.34	&	NSV04126            \\
084304-0342.9	&	0.348563	&	0.3485601	&	3.70	&	3.63	&	4.09	&	3.34	&	-                   \\
093312-8028.5	&	0.406071	&	0.4060657	&	6.59	&	5.22	&	3.11	&	4.97	&	-                   \\
095048-6723.3	&	0.276944	&	0.2769428	&	2.64	&	2.74	&	2.04	&	2.30	&	NSV04657            \\
102552-3224.3	&	0.33706	&	0.3370613	&	2.80	&	2.65	&	2.63	&	2.46	&	-                   \\
114757-6034.0	&	1.65764	&	1.6575598	&	4.25	&	3.59	&	3.11	&	3.16	&	SV~Cen              \\
123244-8726.4	&	0.338519	&	0.3385238	&	4.58	&	8.81	&	7.25	&	6.93	&	NSV~5654            \\
131032-0409.5	&	0.311251	&	0.3112486	&	3.14	&	3.04	&	2.49	&	2.57	&	-                   \\
143103-2417.7	&	0.287859	&	0.2878586	&	3.19	&	2.75	&	2.34	&	2.51	&	-                   \\
144226-4558.1	&	0.251557	&	0.2515636	&	3.29	&	2.75	&	2.20	&	2.69	&	-                   \\
145124-3740.7	&	1.301836	&	1.3017970	&	2.93	&	2.78	&	2.37	&	2.62	&	V0678~Cen           \\
150452-3757.7	&	0.374131	&	0.3741329	&	3.28	&	3.03	&	2.90	&	3.33	&	NSV06917            \\
153152-1541.1	&	0.358259	&	0.3582558	&	2.36	&	2.55	&	3.46	&	3.73	&	VZ~Lib              \\
184644-2736.4	&	0.302836	&	0.3028365	&	3.47	&	2.70	&	2.24	&	2.64	&	-                   \\
195004-5146.7	&	0.87546	&	0.8754432	&	3.57	&	3.55	&	3.44	&	3.90	&	V0343~Tel           \\
195350-5003.5	&	0.286828	&	0.2868259	&	6.96	&	5.85	&	4.40	&	5.92	&	NSV12502            \\
202438-5244.0	&	0.31593	&	0.3159329	&	3.14	&	2.77	&	2.32	&	2.19	&	NP~Tel              \\
213519-2722.8	&	0.3689	&	0.3689034	&	2.76	&	2.60	&	2.18	&	1.89	&	-                   \\
230749-2202.8	&	0.48431	&	0.4843096	&	6.19	&	6.33	&	4.78	&	4.52	&	-                   \\
\hline
\end{tabular}
\caption[LTE search results: ASAS EC objects.]{LTE search results: ASAS EC objects. 
The table lists the object's ID, the period value from the ACVS catalogue, a~corrected 
period obtained during reference model computation. Four $r$ values for 5-, 6-, 7- and 
8-interval runs are given. The last column contains other IDs, if any.}
\label{tab:EC}
\end{table*}

\section{Eclipsing binaries with LTE orbits --- ASAS 123244-8726.4, 
ASAS 075809-4648.5 and ASAS 141035-4546.8}

During a visual inspection of the detected timing variations we identified
three interesting cases most likely corresponding to an LTE effect due
to a third companion. We subsequently analyzed them using a modified
version of our approach. It differs from the original one 
described above in the way the intervals are chosen. Rather than setting the number 
of intervals, this time, the length of the intervals (in days) and the shift between 
them (in days) are used as parameters. This enables one to create overlapping
intervals. E.g. choosing a~length of an interval of 300 days and setting the shift 
between intervals to 100 days produces 30 
intervals assuming that the time span is 3200 days. An important difference is that data 
points are used multiple times for an O--C computation. This way it is possible to produce 
an infinite number of points in the O--C diagram having the same input data as in the 
standard algorithm. The statistical significance of these points is
obviously appropriately lower than in the case of non-overlapping intervals 
and this must be taken into account when deriving formal errors.
The main purpose of such an approach is to investigate how the O--C diagram  
looks between the points calculated using the standard algorithm and also
how this influences the best-fitting LTE orbit. 

%\begin{figure}
%	\includegraphics[angle=-90,width=\columnwidth]{img/AlgorithmDiagram1.ps}
%	\caption{Modified version of the algorithm.}
%	\label{fig:AlgorithmDiagram1}
%\end{figure}

\begin{figure}
\includegraphics[width=\columnwidth]{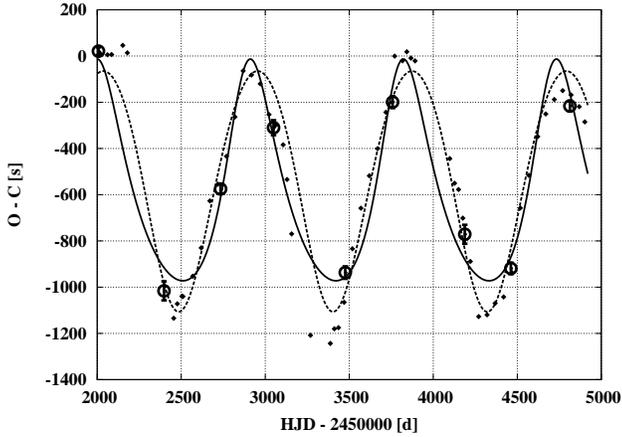}
\caption{ASAS 123244-8726.4. The O--C diagram was obtained using the standard (circles with error bars, solid line) and the overlapping 
methods. Orbital solutions were fitted to both data sets.}
\label{fig:1232all}
\end{figure}

ASAS 123244-8726.4, also known as NSV5654 is an EC binary that has a~period of 0\fd338519. 
It has been identified with high values of $r$ reaching 8.81 in a~6-interval run. The 
O--C plot is shown in Fig. \ref{fig:1232all}. 
A linear trend introduced by an imprecise period was removed by applying a~period correction obtained 
from an orbital fit which included a correction to the period $dP$ as one of the parameters. 
The O--C  diagram calculated using the new period shows evidence of an LTE orbit. 
Three complete cycles seem to be visible. 
The final orbital parameters are summarized in Tab. \ref{tab:07all}.

\begin{table}
\centering
\begin{tabular}{cccc}
\hline
parameter & unit & standard method & overlapping method \\ \hline
\multicolumn{4}{c}{ASAS 123244-8726.4}\\
\hline
$P$ &[d] 		& $911\pm5$ & $918.4$ \\ %\pm 2.4$ \\
$a\sin I$& [AU]	& $0.961 \pm 0.024$ & $1.042$\\ %\pm0.013$\\
$e$ 	&--			& $0.159 \pm 0.019 $ & $ 0.055  $\\ %\pm 0.009$\\
$\omega$ & [deg] & $113\pm8$  & 0.98\\
$T_0$&[HJD] & $2450143\pm17$ & 2450877\\
$f(m_{1,2},m_3)$ & [$M_\odot$] & $0.143\pm0.011$ &0.179 \\
RMS	  &[s] & $67.8$ & -- \\   
$\chi^2$& -- & 3.65 & --\\
$k$ & -- & 3 & -- \\
\hline
\multicolumn{4}{c}{ASAS 075809-4648.5}\\
\hline
$P$ &[d] 		& $1188\pm27$ & $1196$ \\ %\pm 2.4$ \\
$a\sin I$& [AU]	& $0.52 \pm 0.04$ & $0.57$\\ %\pm0.013$\\
$e$ 	&--			& $0.131 \pm 0.052 $ & $ 0.098  $\\ %\pm 0.009$\\
$\omega$ & [deg] & 	$84\pm24 $ & 102\\
$f(m_{1,2},m_3)$ & [$M_\odot$] & $0.0133\pm0.0031$ &0.0173 \\
$T_0$&[HJD] & $2450328\pm80$ & 2450234\\
RMS	  &[s] & $40.77$ & -- \\
$\chi^2$ &-- & 1.43 & --\\
$k$ & -- & 3 & -- \\
\hline
\multicolumn{4}{c}{ASAS 141035-4546.8}\\
\hline
$P$ &[d] 		& $1286\pm120$ & $1165$ \\ %\pm 2.4$ \\
$a\sin I$& [AU]	& $0.76\pm0.13$ & $0.84$\\ %\pm0.013$\\
$e$ 	&--			& $0.0 $ & $ 0.32  $\\ %\pm 0.009$\\
$\omega$ & [deg] & 	$\-- $ & --\\
$T_0$&[HJD] & $2450292\pm122$ &2450154 \\
$f(m_{1,2},m_3)$ & [$M_\odot$] &  $0.035\pm0.019$ & 0.043\\
RMS	  &[s] & $156.15$ & -- \\
$\chi^2$& -- & 2.23 & --\\
$k$ & -- & 2  & -- \\
\hline
\end{tabular}
\caption{LTE orbital parameters: period ($P$), projection of the semimajor axis $a$ on the line of sight ($a\sin I$, $I$ is the orbit's inclination), eccentricity ($e$), argument of periapsis ($\omega$), mass function ($f(m_{1,2},m_3)$) for ASAS 123244-8726.4, 075809-4648.5 and 141035-4546.8 computed using data from the standard 
and the overlapping method. RMS, $\chi^2$ and the degrees of freedom ($k$) have been calculated for the standard method.}
\label{tab:07all}
\end{table}

\begin{figure}
\includegraphics[width=\columnwidth]{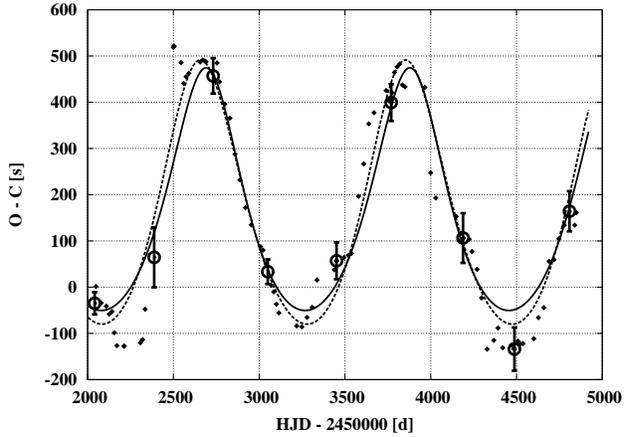}
\caption{ASAS 075809-4648.5. The O--C diagram was obtained using the standard  (circles with error bars, solid line)  and the overlapping 
methods. Orbital solutions were fitted to both data sets.}
\label{fig:07all}
\end{figure}

Another interesting EC object is ASAS 075809-4648.5 or NSV03836. This binary, having a~period 
of 0\fd390383, reveals periodic variability in the O--C diagram. As in the previous case,
we used the standard  (circles with error bars, solid line)  and the overlapping method. Figure \ref{fig:07all} 
shows two sets of O--C data points as well as two corresponding orbital
solutions. Parameters are shown in Tab. \ref{tab:07all}.

\begin{figure}
\includegraphics[width=\columnwidth]{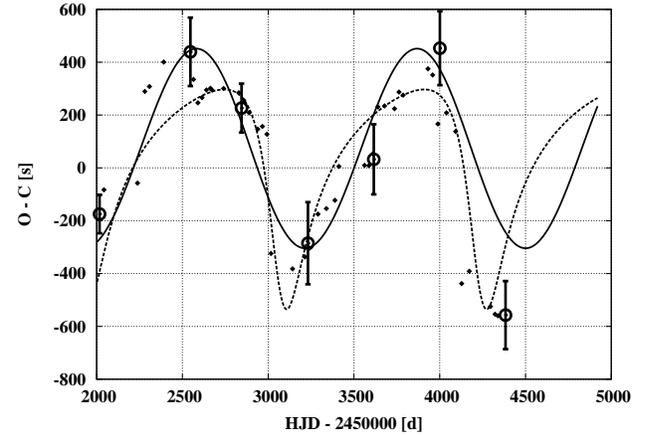}
\caption{ASAS 141035-4546.8. The O--C diagram was obtained using the standard (circles with error bars, solid line) and the overlapping 
methods. Orbital solutions were fitted to both data sets.}
\label{fig:14all}
\end{figure}

ASAS 141035-4546.8 is the only object among analyzed ED systems that  appears to exhibit periodic (O--C) variations with a period clearly shorter than the
data time span. The eclipsing system has a~period of 0\fd988708. 
Figure \ref{fig:14all} shows two sets of the O--C points as well as two
corresponding orbital solutions. In the case of the orbit fitted to the points obtained 
with the standard method, a~circular orbit was assumed. Table \ref{tab:07all} 
provides the orbital parameters.

The above analysis of three interesting cases shows the usefulness of the proposed methods. The 
standard algorithm is well suited for detection and general O--C computation while the second 
method based on overlapping intervals does the interpolation. It is not an interpolation in a~strict 
mathematical sense though. There is no model (linear, cubic, spline, etc.) --
O--C values are calculated accordingly to the actual trend in the data set. As shown in the examples, such 
interpolation can influence the shape of the fitted orbit. The eccentricity is the most sensitive 
parameter. $a\sin I$ and $P$ do not differ much when comparing these two approaches. 

\section{Conclusions}

Eclipsing detached and eclipsing contact binaries were investigated. Altogether 5032 objects have been 
analyzed in terms of the LTE.  Results from 5-, 6-, 7- and 8-interval runs have undergone a~test 
estimating the likelihood of interesting O--C variations. 29 detached and 44 contact binaries have 
passed the final visual tests. Most of the resulting O--C plots have a~parabolic shape meaning a~linear 
period increase or decrease. A few objects reveal 3rd degree variations suggesting a~long period LTE 
orbit, of which only a~short part is visible in the data set. Finally, three diagrams show evidence 
of LTE orbits that have periods shorter than the data's span. These objects have been precisely 
analyzed using a~modified version of the LTE-search algorithm. It uses overlapping intervals and 
generates more points on the O--C plot than the standard approach, thus revealing the possible shape 
of the orbit (without increasing the accuracy). Fitted orbits have semi-major axes smaller than 1AU 
and $\approx3$ year periods.

Obtained O--C plots were compared with known literature data showing compatibility. 

The proposed method is well suited for automated data pipelines due to its versatility. It can handle 
practically any long time-base photometry data and point out most irregularities in eclipse 
timing. 

\textbf{One thing worth mentioning is that, in general, dealing with O--C requires very precise periods.  In our case, the periods came bundled with photometric data from the ASAS Catalogue. If aliases, like $\pm$~10, 20, 30\% of the correct period, shall occur, the algorithm used to detect O--C variations will give a false signal. It must be therefore used with caution and understanding of the process.}

Further simulation work regarding the influence of various light-curve parameters on the 
detection possibility is in progress.  

This work is supported by the Foundation for Polish Science through a 
FOCUS grant and fellowship, by the European Research Council through
the Starting Independent Researcher Grant and Polish MNiSW grant no. N N203 3020 35.

\bsp

\bibliographystyle{mn2e}
\bibliography{bibliography}
\nocite{*}

\label{lastpage}

\end{document}